\documentclass[conference]{IEEEtran}

\usepackage{graphviz}
\usepackage[hidelinks]{hyperref} 

\newcommand{\Autoref}[1]{%
  \begingroup%
  \def\chapterautorefname{Chapter}%
  \def\sectionautorefname{Section}%
  \def\subsectionautorefname{Subsection}%
  \autoref{#1}%
  \endgroup%
}

\usepackage{tikz} 
\newcommand{\encirclebox}[1]{\tikz[baseline={(a.base)}]\node[draw=black,rounded corners=.9ex,inner sep=1pt](a){#1};}

\usepackage[group-separator={,}, group-minimum-digits=4]{siunitx} 
\usepackage{nth} 
\usepackage{pmboxdraw} 
\usepackage{amssymb}
\usepackage{csquotes}
\MakeOuterQuote{"}

\usepackage{etoolbox}
\usepackage{adjustbox} 
\usepackage{amsmath} 
\usepackage[makeroom]{cancel} 
\usepackage{multirow} 

\usepackage{cite}

\ifCLASSINFOpdf
  \graphicspath{{gfx/}{jpeg/}}
\DeclareGraphicsExtensions{.pdf,.jpeg,.png}
\else

\fi

\usepackage{url}

\usepackage{tcolorbox}

\begin{document}
\newcommand{\toniot}{TON\_{}IoT}
\newcommand{\citeMISSING}[1][]{\,{\color{red}[!ref!]}}
\newcommand{\UNFINISHED}{\,{\color{red}[UNFINISHED]}}
%
\title{Privacy-Preserving Intrusion Detection using Convolutional Neural Networks}
%
%


\author{\IEEEauthorblockN{Martin Kodyš}
\IEEEauthorblockA{
\textit{ST Engineering}\\
Singapore \\
}
\and
\IEEEauthorblockN{Zhongmin Dai}
\IEEEauthorblockA{
\textit{ST Engineering}\\
Singapore \\
}
\and
\IEEEauthorblockN{Vrizlynn L. L. Thing}
\IEEEauthorblockA{
\textit{ST Engineering}\\
Singapore \\
}
}


%


\maketitle


\begin{abstract}

Privacy-preserving analytics is designed to protect valuable assets. A common service provision involves the input data from the client and the model on the analyst's side. The importance of the privacy preservation is fuelled by legal obligations and intellectual property concerns.
We explore the use case of a model owner providing an analytic service on customer's private data. No information about the data shall be revealed to the analyst and no information about the model shall be leaked to the customer.
Current methods involve costs: accuracy deterioration and computational complexity. The complexity, in turn, results in a longer processing time, increased requirement on computing resources, and involves data communication between the client and the server.
In order to deploy such service architecture, we need to evaluate the optimal setting that fits the constraints. And that is what this paper addresses.
In this work, we enhance an attack detection system based on Convolutional Neural Networks with privacy-preserving technology based on PriMIA framework that is initially designed for medical data.

\end{abstract}


\section{Introduction}

Privacy-preserving technologies gain popularity as the value of both, the know how and the data, increases. One of the areas of interest is Machine Learning where protection is required in multiple points: data used for training, model created from the data (both, the intellectual property and the traces of the training data), and data used for inference. 

In the description of such setting, we can discern three \emph{roles}: Training Data Owner (TDO), Model Owner (MO), Inference Data Owner (IDO).

In case of a provision of the model as a service, the Model Owner must ensure the protection of the assets of Data Owners against leaks, as well as defend their own model against exfiltration.

The intrusion detection task \cite{kodys2021intrusion} using Convolutional Neural Networks provides a basis for this work. This basis is then augmented with privacy-preservation techniques following the PriMIA framework \cite{Kaissis2021PriMIA}.

We point out important factors for the optimisation of hyper-parameters, especially the fixed fractional precision that steers the accuracy of the network.

The paper is structured as follows:
\Autoref{sec:background} provides background in privacy preservation and Intrusion Detection on IoT data.
\Autoref{sec:related_works} details current solutions and compare them to the work presented in this paper.
\Autoref{sec:design} explains the design of the solution. 
\Autoref{sec:experiments} summarises the experiments and the obtained results, followed by \autoref{sec:discussion} that offers a discussion on possible evolutions and applications, thereafter concluded in \autoref{sec:conclusion}.

\section{Background} 
\label{sec:background}

Two main themes intersect in this paper: intrusion detection in the Internet of Things (IoT) based on the actual data readings and privacy-preserving technologies. We present the perspective of an privacy-concious intrusion detection service provider. Our goal is to provide a reliable service without being aware of the actual client's data and without revealing our model to the client.

\subsection{Intrusion Detection in IoT} 
\label{sub:intrusion_detection_in_iot}
The task of intrusion detection can be formulated as a prediction of a system status class based on a set of observations up to the moment.

In our case, we analyse the data coming from an IoT system as described in \cite{kodys2021intrusion}. The implemented technique is using Convolutional Neural Networks architecture. The input data consists of pre-processed sensor readings in the form of a 2-dimensional picture, a slit-view of chronological sensor readings. The output is one of the 8 classes -- a benign state \texttt{normal}, or one of 7 different attacks: \texttt{backdoor},  \texttt{ddos}, \texttt{injection}, \texttt{password}, \texttt{ransomware}, \texttt{scanning}, or \texttt{xss}. These classes are defined in the \toniot{} "TrainTest" dataset\cite{alsaedi2020ton_iot}. The TrainTest was pre-processed according to \texttt{TT500n} aggregation strategy described in \cite{kodys2021intrusion}. This means that \texttt{TT} (standing for "TrainTest") was partitioned in sequences of \texttt{500} readings, with \texttt{n}o aggregation by time. To conserve temporal patterns, each 500-item long sequence as a whole went into train or test set. Within their respective sets, these sequences were concatenated amd a sliding window produced the input data according to our target architecture, i.e., 224 readings.

\begin{figure}[!htbp]
  \includegraphics[width=\columnwidth]{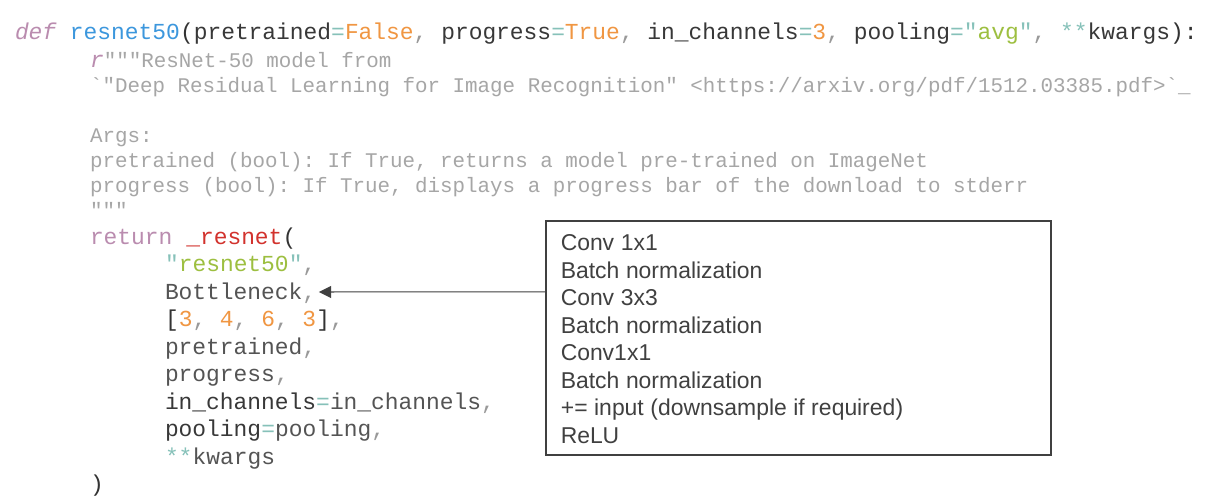}

  \caption{Extension of PriMIA to implement ResNet50. Bottleneck had already been defined as specified in the inner frame. This implementation corresponds to \cite{he2016resnet}. The array [3, 4, 6, 3] defines the number of successive Bottleneck blocks before a change of dimensions.}
  \label{fig:primia-resnet50-implementation}
\end{figure}

The architecture selected for privacy-preserving analytics was \emph{ResNet50}\cite{he2016resnet}. In the attack classification task, the performance of \emph{ResNet50} and \emph{EfficientNet-B0} was relatively close. Between them, we chose pragmatically the one sharing building blocks with the already implemented \emph{ResNet18} architecture within PriMIA framework\cite{Kaissis2021PriMIA}. The required modification is shown in \autoref{fig:primia-resnet50-implementation}.

\begin{figure}[!htbp]
  \includegraphics[width=\columnwidth]{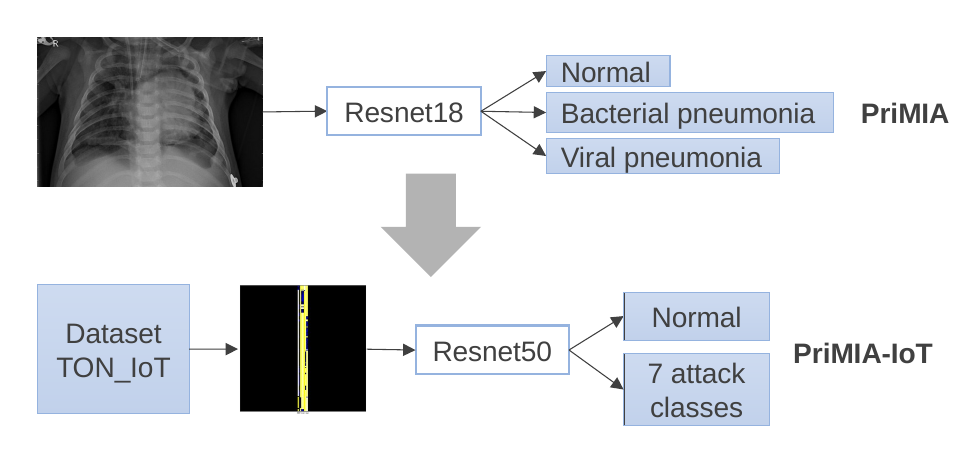}

  \caption{Adjustments of PriMIA to process data from IoT detection use case.}
  \label{fig:primia-iot}
\end{figure}

The adjustment of the PriMIA framework to our use case is illustrated in \autoref{fig:primia-iot}. In simple terms, the \toniot{} dataset was converted into RGB images. The hard-coded 1-channel greyscale image input was extended to allow for 3-channel inputs. We completed the implementation of ResNet50 that reused the blocks available in ResNet18. And finally, the number of output classes was extended from 3 to 8. More details are shared in \autoref{sec:design}.

The intrusion detection technique from \cite{kodys2021intrusion} relies on the 2-dimensional representation of the sensor readings in time, which is perceived as an image corresponding to a table or a grid. The position on the x-axis determines different IoT (sub)sensors and the position on y-axis determines the point in time, with on top the oldest within the given window and featuring the most recent readings on the last row of the image or table.

Given 17 subsensor outputs, there are 17 columns placed in the middle and padded with value \texttt{0} on the left and right to fit 224 columns (translated to pixels). As not all the sensors provide outputs within the same period of time (e.g., 1 second), several imputation strategies were explored in \cite{kodys2021intrusion}.

\begin{figure}[!htbp]
  \centering
  \includegraphics[width=.5\columnwidth]{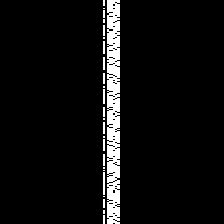}
  \centering
  \caption{Example of an encoded $224\times224\times3\text{-channel}$ picture representing a 224 items long history (vertically) of 17 different subsensors (horizontally) as a sensor provides one or more subsensor readings.}
  \label{fig:a-sample-password-miss3}
\end{figure}

In this work, we focused on a simple and radical strategy \texttt{miss3} that replaces 
actual values by \texttt{0} and missing ones by value \texttt{1}. As indicated by number "3" in its name, the method is used to fill all three channels equally to obtain an RGB picture. An example of a resulting picture after normalisation is presented in \autoref{fig:a-sample-password-miss3}. 

\subsection{Privacy-Preserving Technologies} 
\label{sub:privacy_preserving_technologies}

Many ways are being explored\cite{domingo2020ppt} in search of an ideal privacy-preserving technology, with negligible overhead and optimal security adapted to requirements. The research falls into the field of Secure Multi-Party Computation (MPC). And from our standpoint, we will focus on protecting the assets against other parties that may be part of the computation.

Although the Homomorphic Encryption is a promising technology that is being actively researched to achieve high performance \cite{jung2021gpu100xCKKS}, Fully Homomorphic Encryption solutions still incur an unaffordable computational cost, especially in the case of large neural network architectures. Hybrid approaches delegating expensive tasks to the Data Owner\cite{Tian2022sphinx} generate high communication costs, undermining the relevance of the server side altogether.


A common practice is to defend against the adversary that is \emph{honest but curious} (also called \emph{semi-honest} in \cite{oded2009foundations}). It is the case of PriMIA where the main use case is collaboration of several medical establishments to produce a common model and then to share it safely with a non-participating medical establishment to apply the know-how but without giving it away. An honest-but-curious adversary does not attempt to break the protocol, although they might extract any information from the available data. In practice, the common model shared among all participating medical establishments could be subjected to an inversion attack. To prevent it, PriMIA employs an optional Differential Privacy technique.

Privacy preservation is often evaluated as indistinguishability of the intercepted data from a random strings. The schemes capturing the security guarantees continue evolving, including IND-CPA-D\cite{Li_Micciancio_2021_ind-cpa-d}, i.e., \emph{indistinguishability under chosen plaintext attacks with decryption oracles}, that builds on the notion of indistinguishability under chosen plaintext attacks (IND-CPA) and adds the possibility for the attacker to interact with a restricted oracle. This notion corresponds to the cases of passive attacks to which approximative arithmetic homomorphic schemes like CKKS \cite{ckks2017} were vulnerable.

The FSS paper\cite{Boyle2019fss} mentions the query quota as the only mitigation for model inversion attacks to limit the amount of information the attacker might obtain to reconstruct the model. The more complex neural network, the slower the evaluation, which naturally, provides a quota-like protection.

We can distinguish the MLaaS with training and without training: We situate our research in the case where we have an already trained model to which we want to offer a secure access without sharing it. In our case, the training is excluded from the problem this paper addresses.

In different cases, the service may be Cloud-like\cite{Tian2022sphinx}, where the service provider would give access to their infrastructure and guarantee the privacy of its usage. The client would privately input data for the training, decide an architecture, train it, and provide inference -- all without letting the external observer or service provider learn anything about the data and the model. This described mode of operation is suitable for infrastructure service providers (e.g., Cloud) and may find a practical use in systems relying on a public blockchain. However, we will focus on providing a private access to the inference service. The Model Owner will not have any access to the data provided by the Inference Data Owner. And vice-versa, the Inference Data Owner will not be able to reconstruct the model and thus protecting not only the intellectual property of the Model Owner but also the Training Data Owners against inversion attacks.


\section{Related Works}
\label{sec:related_works}

Concerning our specific use-case of Machine Learning as a Service (MLaaS), we discern several aspects that have been addressed in the literature. Generally, privacy preserving technologies protect data in different stages of the computation but also the computation process itself. We will focus on technologies that do not rely on trust amongst different parties that are part of the computation as defined in \autoref{sec:background}.

Machine Learning, and in particular, Deep Learning were objects of privacy-preservation efforts. The most straightforward way to protect the Training Data Owner is the use of techniques of input data perturbation while conserving useful features or properties. In particular, the concept of Differential Privacy (DP)\cite{abadi2016dp} is becoming popular as the perturbation is quantified and provides a framework for expressing the privacy cost in terms of accuracy. DP can improve privacy for Training Data Owner while producing a model, where it effectively hides the details from the Model Owner. However, it does not provide any advantage to Inference Data Owner, as the predictions must not be biased.

In order to protect the IDO's assets, computation using homomorphic encryption is gaining interest. However, the current solutions are mostly relevant for simpler Machine Learning algorithms. For example, CryptoNets\cite{dowlin2016cryptonets} operate on MNIST with dimensions of $28\times28$ pixels and uses client-side evaluation of activation functions.

Falcon\cite{wagh2021falcon} offers security against a malicious adversary in 3-party setting. However, it requires an honest majority. This is an issue that does not have a simple solution if only two parties are to be involved. Any third party affiliated to any of the two would present a risk of collusion and, thus, breaking down the trust in the system.

\section{Design} 
\label{sec:design}
The PriMIA framework extends PySyft framework of its version 0.2.9, in particular its PyGrid component for remote execution of Machine Learning tasks. We adapt the solution of intrusion detection \cite{kodys2021intrusion} to be compatible with PriMIA.

An implementation of ResNet50 is effortless as the basic building blocks were ready in PriMIA's implementation of ResNet18.

\subsection{PriMIA framework} 
\label{sub:primia_framework}
Used as a basis for our contribution, we focus on specific aspects of the PriMIA framework. The framework is open source, is interpreted in Python 3.7, and requires CUDA 10.x to enable GPU acceleration for training although in our experimentation, the GPU was not involved during inference. In this paper we will look closer at the inference process.

PriMIA's input data has a format of $224 \times 224 \times 1$ tensor, i.e., a 1-channel (usually representing a greyscale) square image with the dimensions $224 \times 224$.

\begin{figure*}[!htbp]
  \includegraphics[width=\textwidth]{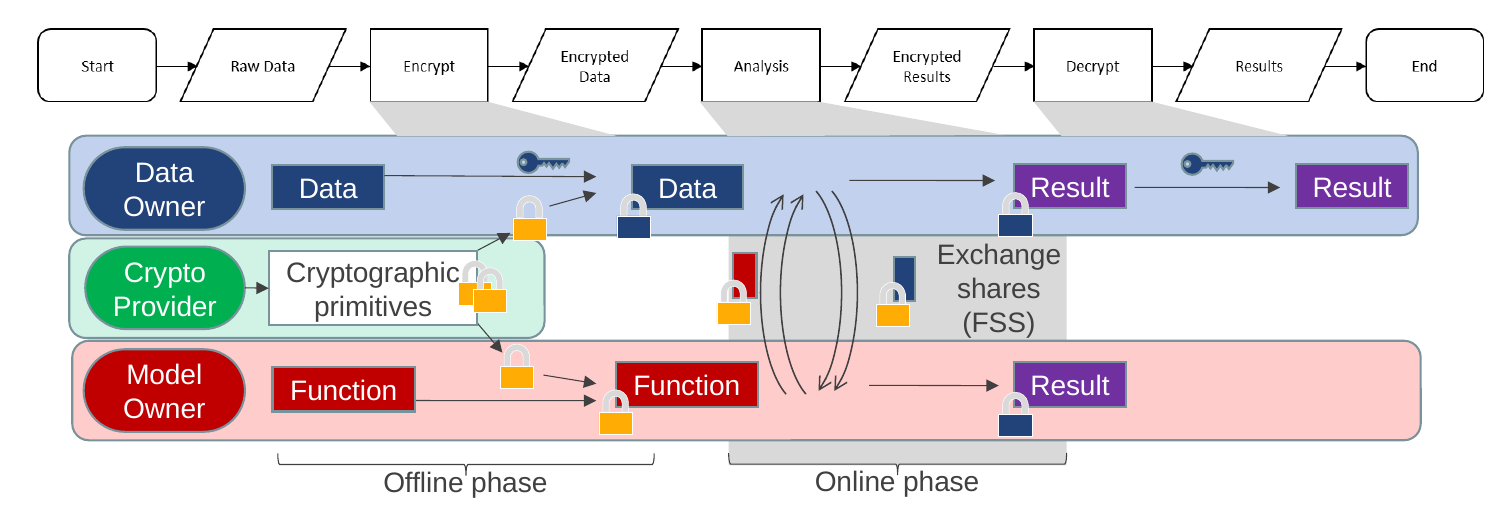}

  \caption{Mapping of PriMIA framework to the use case of privacy-preserving analytics}
  \label{fig:primia}
\end{figure*}

The inference scenario is depicted in \autoref{fig:primia} relying on \emph{Function Secret Sharing (FSS)}\cite{Boyle2019fss}. The client prepares their raw data, encrypts it using the trusted source of correlated randomness. Meanwhile, the server prepares shares of the function that computes the model. The masked shares are then exchanged to obtain the encrypted result. The client then uses their secret key to get the cleartext result.

The correlated randomness is particularly important in the Secure Multi-Party Computation. PriMIA falls into the category of MPC with pre-processing according to definitions in \cite{ishai2013randomness}. This means that a speed-up is achieved by generating randomness in parallel, independently of actual encryptions.

\subsection{Fixed Fractional Precision} 
\label{sub:fixed_fractional_precision}

What enables the possibility of shared secret computation, are the properties of integers that can be masked with the random strings to obtain additive secret shares. Not all representations of numbers allow the exploitation of this property. In particular, a common floating point implementation breaks these properties. The original PriMIA implementation uses \emph{fixed fractional precision} parameter to handle the conversion between a decimal value and its expression in memory as an integer with a given length, multiples of a pre-defined factor. This conserves additive share properties equally for very large and small values. 

PriMIA, through the underlying PySyft\footnote{PySyft v0.2.9: \url{https://github.com/OpenMined/PySyft/tree/syft_0.2.x}}, uses integers 64 or 32 bits long. An integer $m$ represents a decimal value $z$, following the equation: $m = z \times {b^{p}}$, where base $b=10$ and precision $p \in [\![1,16]\!]$. Therefore, this system can represent decimal numbers up to $p$ decimal places within the interval $[-\frac{2^{s-1}}{b^p}; \frac{2^{s-1}-1}{b^p}]$, which for \texttt{long} of size $s=64$, base $b=10$, and precision $p=4$ provides an approximative range of $[-9.22\times10^{14}; 9.22\times10^{14}]$, while a precision $p=16$ represents only values between $-922$ and $922$.

These approximative ranges provide an insight why a computation with high (or low) precision might fail - due to a value overflow (or underflow). This is especially true for Machine Learning where the gradients risk being exploding (or vanishing). The unwanted overflow and underflow interfere with additive shared secrets as well, where an "overflow" is part of the design of the encryption/decryption as the addition is performed given a specific modulo.

\begin{figure}[!htbp]
  \includegraphics[width=\columnwidth]{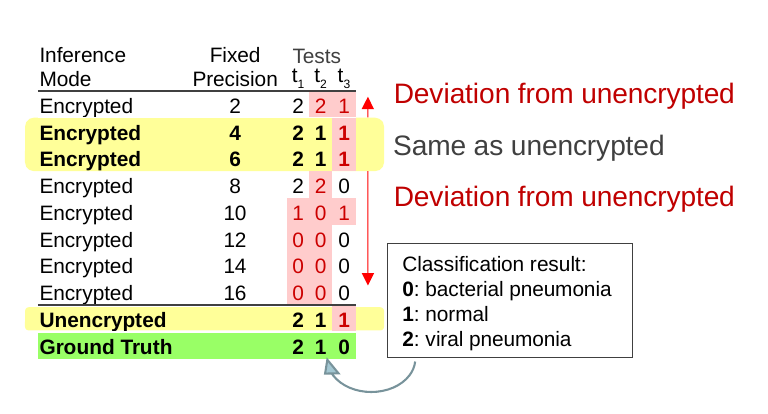}

  \caption{Fixed fractional precision hyperparameter search. In the original PriMIA setting, there are three classes (0, 1, 2). Changing the fixed fractional precision changes how well the encrypted inference matches the unencrypted inference results. This table provides evidence of setting up of the precision is crucial to get encrypted inference function properly.}
  \label{fig:fixed-fractional-precision}
\end{figure}

In order to achieve the optimal results (best matching to the original results on non-encrypted model), we perform an exhaustive search on the hyperparameter of the \emph{fixed fractional precision}. The results are presented in \autoref{fig:fixed-fractional-precision}. Note that each of the three tests ($t_1$, $t_2$, $t_3$) was a single-item inference whose ground truth is presented in the bottom of the table. Even though we can see a clear pattern of which values of fixed fractional precision are optimal for this network and dataset, it might not be the case for all networks and all data.

\subsection{Function Secret Sharing} 
\label{sub:function_secret_sharing}
PriMIA's code\footnote{\url{https://github.com/gkaissis/PriMIA}} claims that there are differences between the Function Secret Sharing defined in the paper\cite{Boyle2019fss}, although they are not specified and a detailed analysis might be required to establish the differences.

The FSS implementation in PriMIA requires three parties, from which only two are needed when dealing with the actual inference. The third party, a crypto-provider is the source of the correlated randomness. It is used to generate masking keys for the exchanges between the two active parties: the Inference Data Owner (\emph{data owner} in PriMIA), and the Model Owner (also \emph{model owner} in PriMIA).

The original FSS scheme and the improvements\cite{Boyle2019fss} were evaluated under the assumption of a semi-honest\cite{oded2009foundations} adversary that is honest but curious. 
That is, an adversary that has the interest to follow the protocol but is interested in exploiting possible leaks to gain an insight in the other party's data. In order to  defend against a malicious adversary, the authors suggest an extension by simple authentication methods.

The FSS is also \emph{perfect secret sharing} scheme that does not leak any information about the plaintext from the ciphertext. The encryption is based on the principle of a one-time pad. Additive secret sharing protocol generates correlated randomness that serves to hide the plaintext.

\section{Experiments} 
\label{sec:experiments}

Experiments were conducted to assess the overhead of the solution compared to an unencrypted computation. The inference stage is the focus of our use case of Machine Learning as a Service.

We monitored the time, memory resources, and exactness with respect to the unencrypted inference.

\subsection{Configuration} 
\label{sub:configuration}
The results were achieved using two configurations $M_1$ for GPU-enabled training and $M_2$ for inference task:
\begin{itemize}
	\item Machine $M_1$: 
Ubuntu 20.04.1 LTS focal: Linux 5.4.0-89-generic \#100-Ubuntu SMP Fri Sep 24 14:50:10 UTC 2021 x86\_64 GNU/Linux
64 GB RAM - \textbf{32 cores} - Intel® Xeon® Silver 4215 CPU @ \textbf{2.50Ghz} - Nvidia-driver 470.161.03; CUDA 10.1.243
	\item 	Machine $M_2$:
Ubuntu 22.04.1 LTS jammy: Linux 5.15.0-56-generic \#62-Ubuntu SMP Tue Nov 22 19:54:14 UTC 2022 x86\_64 GNU/Linux
64 GB RAM - \textbf{16 cores} - Intel® Core™ i7-10700K CPU @ \textbf{3.80GHz} - GPU not in use in the inference task
\end{itemize}

\subsection{Methodology} 
\label{sub:methodology}
The departure point was to reproduce the results from \cite{kodys2021intrusion} using Deep Learning framework of PyTorch instead of Keras models that were used in the aforementioned work. ResNet50 model pre-trained on ImageNet was used for the initial training on $M_1$: 20 epochs on 14,995 images took 1\,h 07\,min. 

\begin{figure}[!htbp]
	\centering
	\includegraphics[width=\columnwidth]{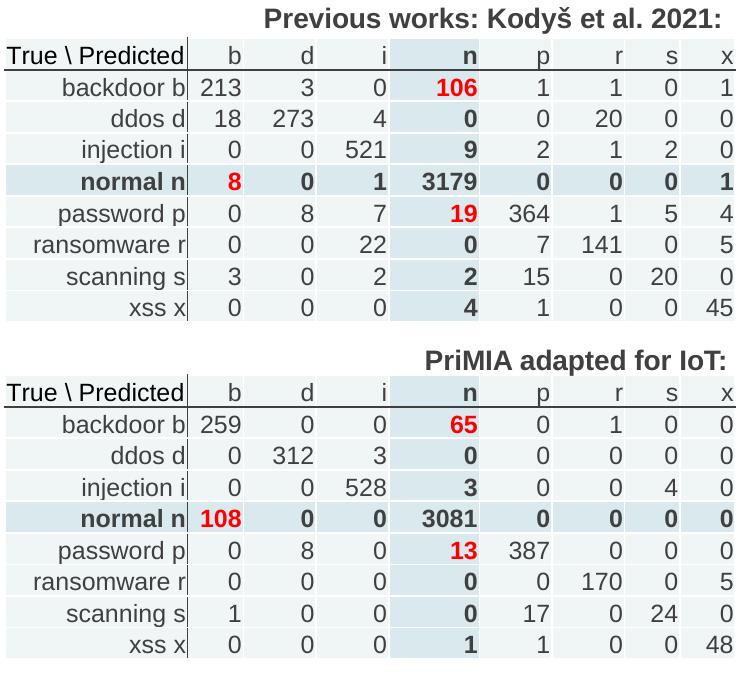}
	\caption{Confusion matrices of two models: upper table is from \cite{kodys2021intrusion}; lower table represents this work.\\
  Both inferences were performed on \toniot{} dataset \texttt{TT500n}, with \texttt{miss3} imputation strategy that fills in missing values with \texttt{1} and actual values are replaced with \texttt{0}.\\
	More significant differences between \emph{normal} and \emph{attack} classes are highlighted in bold red font.}
	\label{fig:compare-unencrypted}
\end{figure}

Its performance is similar to the original work\cite{kodys2021intrusion}. Their comparison is available in \autoref{tab:models-binary-metrics} and their respective confusion matrices in \autoref{fig:compare-unencrypted}. The differences between the inference results of the two Deep Learning frameworks are sufficiently small to be equivalent for the purpose of this study.

The reason to consider a 5\% subset is to obtain results of the encrypted inference within a reasonable time, given that the inference on the full dataset would take approximatively 42 days.

We possess an unencrypted model that is the property of the Model Owner and will be protected by the applied technology according to our use case.

The performance of encrypted inference is measured on $M_2$, and compared to unencrypted inference. No GPU accelerated implementation of FSS was available as of June 2023.

\subsection{Results} 
\label{sub:results}
The presented results were obtained on ResNet50 architecture implemented in PyTorch. We chose to use \texttt{TT500n} dataset for comparison to the original \toniot{} sensor readings dataset. The imputation strategy \texttt{miss3} was chosen for its simplicity and results on par with more complex imputation strategies.
Input for our ResNet50 architecture is an image $224\times224\times3$.

\Autoref{tab:models-binary-metrics} compares the binary classification metrics for the reference model trained in \cite{kodys2021intrusion} and our reimplementation. Additionally, we test the same model on a subset of 5\% of the  \texttt{TT500n\_{}miss3} dataset. We note that an incidental improvement of FPR to 0\% and a deterioration of TPR are visible but irrelevant - as they are due to an implicit bias of the random sample.

The comparison of unencrypted inference and its encrypted versions are shown in \autoref{tab:performance-encrypted}, and further details of the inference on 251 items of the 5\% \texttt{TT500n\_{}miss3} dataset in form of confusion matrices \autoref{tab:confusion-matrix-251-unencrypted} and \autoref{tab:confusion-matrix-251-encrypted}. We verified item by item that the encircled values are the only change in predictions: a single was classified as \texttt{ransomware} (an \emph{attack} class) instead of originally correct \texttt{backdoor} (also an attack class in the binary classification). We note that the binary classification metrics for unencrypted and encrypted inference have the exact same values and correspond to those reported in the last column of \autoref{tab:models-binary-metrics}. The error in encrypted inference does not translate into any change of binary classification metrics as it remains within the set of \emph{attack} classes. 
\begin{table}[!t]
  \caption{Binary Classification Metrics}
  \label{tab:models-binary-metrics}
  \centering
  \begin{tabular}{|r|S[table-format=3.1]|S[table-format=3.1]|S[table-format=3.1]|}
  \cline{2-4}
  \multicolumn{1}{c|}{}& {\cite{kodys2021intrusion}: Keras} & \multicolumn{2}{c|}{This paper: PyTorch} \\
  \hline
  Dataset     & {TT500n\_{}miss3}       &  {TT500n\_{}miss3}  & {\begin{tabular}{@{}c@{}}5\%\,TT500n\_{}miss3\end{tabular}$^{\mathrm{*}}$} \\

  Cardinality &\num{5039} & \num{5039}          & \num{251} \\      
  \hline
  Accuracy    &\num{94.4}{\%} & \num{96.2}{\%}  & \num{97.6}{\%} \\ 
  TPR         &\num{92.4}{\%} & \num{95.6}{\%}  & \num{93.5}{\%} \\ 
  FPR         & \num{0.3}{\%} &  \num{3.4}{\%}  & \num{0}{\%}   \\  
  \hline

  \multicolumn{4}{p{.9\columnwidth}}{\rule{0pt}{1.5em}$^{\mathrm{*}}$ Both, encrypted and unencrypted cases yield the same values.}
  \end{tabular}
\end{table}

\begin{table}[!t]
  \caption{Duration and Matching of Encrypted Inference}
  \label{tab:performance-encrypted}
  \centering
  \begin{tabular}{|r|c|c|c|c|}
  \cline{2-4}
    \multicolumn{1}{c|}{} & {Unencrypted} & \multicolumn{2}{c|}{Encrypted}  \\
    \multicolumn{1}{c|}{} &  CPU          & Local         & HTTP            \\
  \hline
    {\begin{tabular}{@{}r@{}}Inference duration \\
     per item\end{tabular}}                        & \num{0.15}\,s   &  {\begin{tabular}{@{}c@{}}\num{677.31}\,s \\ { = 11\,min\,17\,s}\\ \end{tabular}}  & {\begin{tabular}{@{}c@{}}\num{1356.15}\,s \\ { = 22\,min\,36\,s}\end{tabular}}   \\
    \hline
    Matching              & \num{100}{\%} & {\begin{tabular}{@{}c@{}}\num{99.6}{\% } \\ { = }\num{250}{/}\num{251}\end{tabular}} & {$^{\mathrm{*}}$}             \\
  \hline

  \multicolumn{4}{p{.9\columnwidth}}{\rule{0pt}{1.5em}$^{\mathrm{*}}$ Same as the Local Encrypted computation because only network stack differs, the computation algorithm is the same.}
  \end{tabular}
\end{table}

\begin{table}
  \caption{Unencrypted inference -- Confusion matrix}
  \label{tab:confusion-matrix-251-unencrypted}
  \centering
  \begin{tabular}{r @{ } c|r r r r r r r r}
    \multicolumn{2}{c|}{True Label}
          & \multicolumn{8}{c}{Predicted Label} \\
          &   & b & d & i & n & p & r & s & x \\
    \hline
    backdoor  & b & \encirclebox{11} & 0 & 0 & 5 & 0 & \encirclebox{0} & 0 & 0  \\
    ddos      & d & 0 & 15 & 0 & 0 & 0 & 0 & 0 & 0  \\
    injection & i & 0 & 0 & 25 & 0 & 0 & 0 & 1 & 0  \\
    normal    & n & 0 & 0 & 0 & 159 & 0 & 0 & 0 & 0  \\
    password  & p & 0 & 0 & 0 & 1 & 19 & 0 & 0 & 0  \\
    ransomware& r & 0 & 0 & 0 & 0 & 0 & 9 & 0 & 0  \\
    scanning  & s & 0 & 0 & 0 & 0 & 2 & 0 & 1 & 0  \\
    xss       & x & 0 & 0 & 0 & 0 & 0 & 0 & 0 & 3  \\
  \end{tabular}
\end{table}

\begin{table}
  \caption{Encrypted inference -- Confusion matrix}
  \label{tab:confusion-matrix-251-encrypted}
  \centering
  \begin{tabular}{r @{ } c|r r r r r r r r}
    \multicolumn{2}{c|}{True Label}
          & \multicolumn{8}{c}{Predicted Label} \\
          &   & b & d & i & n & p & r & s & x \\
    \hline
    backdoor  & b & \encirclebox{10} & 0 & 0 & 5 & 0 & \encirclebox{1} & 0 & 0  \\
    ddos      & d & 0 & 15 & 0 & 0 & 0 & 0 & 0 & 0  \\
    injection & i & 0 & 0 & 25 & 0 & 0 & 0 & 1 & 0  \\
    normal    & n & 0 & 0 & 0 & 159 & 0 & 0 & 0 & 0  \\
    password  & p & 0 & 0 & 0 & 1 & 19 & 0 & 0 & 0  \\
    ransomware& r & 0 & 0 & 0 & 0 & 0 & 9 & 0 & 0  \\
    scanning  & s & 0 & 0 & 0 & 0 & 2 & 0 & 1 & 0  \\
    xss       & x & 0 & 0 & 0 & 0 & 0 & 0 & 0 & 3  \\
  \end{tabular}
\end{table}


\section{Discussion} 
\label{sec:discussion}
During the experimentation with PriMIA framework, the key parameter to have an immense influence was the \emph{fixed fractional precision}. Its adjustment decided whether the resulting output ended up the same value or it followed the unencrypted model. Either underflow or overflow degraded the performance. An extension of this work can focus on automation of the optimisation of this process beyond the naive method of enumeration search that we used in this paper. Further research can lead to a more comprehensive optimisation techniques.

Further research can be conducted to compare this approach to other privacy-preserving technologies. Especially, Fully Homomorphic Encryption shows substantial improvements in recent years and the efforts to improve the efficiency using GPU are under way.

Other Secure Multi-Party Computation protocols (e.g., Falcon\cite{wagh2021falcon}) are secure against a minority of malicious parties. Their use in 2-party setting is problematic because both parties are \emph{de facto} required to be honest. It is also to be noted that the FSS implementation has an implicit third party, a trusted dealer -- the source of correlated randomness. The original paper on secure computation via FSS\cite{Boyle2019fss} mentions the possibility of extension to malicious security model referring to authentication techniques of \cite{bendlin2011she_mpc,ishai2013power_randomness_SC}, the issue can be mitigated by including tests in the protocol and abort if the tests show a deviation from the protocol. More insights about correlated randomness, on which many protocols rely, can be found in \cite{ishai2013power_randomness_SC}.

\section{Conclusion} 
\label{sec:conclusion}
This work applies privacy-preserving technology designed for medical imaging to intrusion detection.

Through the use case of providing a secured service of inference on customer's private data and ensuring the privacy of our own model, we illustrated different issues and possible ways forward to solve them.

Following the current state of the art, we show the results of an application to ResNet50 allow for an inference through secure channel. The current speed of the inference is suitable for demonstration purposes using private data rather than a deployable Machine Learning as a Service solution.

\bibliographystyle{IEEEtran}
\bibliography{main}

\end{document}